\documentstyle[prd,aps,floats,epsf]{revtex}
\newcommand{\be}{\begin{equation}}
\newcommand{\ee}{\end{equation}}

\begin{document}
\preprint{PURD-TH-0001, hep-ph/yymmddd}
\draft

\renewcommand{\topfraction}{0.99}
\renewcommand{\bottomfraction}{0.99}
\twocolumn[\hsize\textwidth\columnwidth\hsize\csname
@twocolumnfalse\endcsname

\title
{\Large {\bf Resonant Amplification of Gauge Fields in Expanding
Universe
%Electromagnetic Fields
}}
\author{Fabio Finelli$^{\, 1,2,3}$ 
and Alessandro Gruppuso$^{\, 2}$} 
\address{
$^1$ Department of Physics, Purdue University, West Lafayette, IN 47907,
USA \\
$^2$Dipartimento di Fisica, Universit\`a degli Studi di Bologna
and I.N.F.N., \\ via Irnerio, 46 -- 40126 Bologna -- Italy \\
$^3$Istituto Te.S.R.E./CNR, via Gobetti
101 -- 40129
Bologna -- Italy}
\date{\today}
\maketitle
\begin{abstract}
We investigate the possibility that gauge fields 
%electromagnetic fluctuations 1
are amplified in an expanding universe by parametric resonance, during the
oscillatory regime of a scalar field to which they are coupled. 
We investigate the coupling of gauge fields to a charged scalar field and
to an axion.
For both couplings, gauge fields fluctuations undergo 
exponential instabilities. We discuss how the presence of other charges 
or currents may counteract
the resonance, but we argue that in some cases the resonance will
persist and that hence this mechanism could have some
relevance for the problem of large scale primordial
magnetic fields.
\end{abstract}

]

\vskip 0.4cm
 
\section{Introduction} 
In these last years the concept of parametric
resonance (henceforth PR) has played a 
fundamental role in cosmology for the development of the theory of
preheating \cite{tutti}, the explosive decay of the inflaton after
inflation. There have been investigations of several types of fields
coupled to the inflaton: scalars \cite{tutti}, fermions \cite{fermion}, 
gravitational waves \cite{gw}, scalar gravitational perturbations 
\cite{scalar}, gravitinos \cite{gravit}.

Systems which includes {\em nearly} \cite{nearly} 
conformal invariant fields can be efficiently amplified in
an expanding universe \cite{greene}. This occurs because the 
Robertson-Walker metric is conformally related to the flat metric,  
and therefore the equations of motion can be mapped 
into similar ones in Minkowski space-time.
In these systems the resonance can be efficient without resorting to large
couplings \cite{klsbig} (it is sufficient to have $q \sim {\cal O} (1)$,
where $q$ is the
amplitude of the driving term). The resonance is not weakened by the
expansion of the universe and it 
persists until nonlinear effects will shut off it \cite{greene}.

It is interesting to investigate the effect of PR on fields which are 
conformally coupled to gravity. 
For these fields particle production due
to the expansion of the universe is absent \cite{conformal}. Therefore PR
can be an alternative and efficient way to amplify fluctuations of these
fields. 
%which are 
%conformally coupled to gravity. 
Examples of conformally invariant fields are massless
vector fields and massless fermions \cite{conformal}. 
In the case of fermion the Pauli blocking
is present, but this does not prevent the occurring of novel
interesting effects \cite{fermion}. For a massless vector field 
an exponential Bose enhancement occurs. In this Letter we investigate 
the evolution of gauge field coupled to a charged scalar field and to an
axion. 

\section{Scalar Electrodynamics} 

The Lagrangian density
for scalar electrodynamics is given by 
\be
{\cal L} = - \frac{1}{16 \pi} F_{\mu \nu} F^{\mu \nu} - (D_\mu \Phi)^*
(D^\mu \Phi) - V(\Phi^* \Phi)
\label{scalarlag}
\ee     
where $D_\mu = \nabla_\mu - i e A_\mu$ 
is the gauge and metric covariant derivative,  
$A_\mu$ is the gauge potential and 
$F_{\mu \nu}$ is the antysimmetric field tensor, defined as $F_{\mu
\nu} = \nabla_\mu A_\nu - \nabla_\nu A_\mu$.
From this Lagrangian the equations of motion are:
\begin{eqnarray}
- \Box \Phi + 2 i e A_\mu \partial^\mu \Phi + i e \Phi \nabla_\mu A^\mu 
+ \frac{\partial V}{\partial \Phi^*} + e^2 A_\mu A^\mu \Phi = 0 
\label{scafield} \\
\nabla_\nu F^{\nu \mu} = - 4 \pi j^\mu + 8 \pi e^2 A^\mu |\Phi|^2
\label{emfield}
\end{eqnarray}
where $j_\mu = i e (\Phi \partial_\mu \Phi^* - 
\Phi^* \partial_\mu \Phi)$. 
We shall consider inhomogeneous linearized fluctuations
around homogeneous quantities $f (t, {\bf x}) = {\bar f} (t) + \delta
f (t, {\bf x})$ (where $f$ is any field)
%$\Phi(t, {\bf x}) = {\bar \Phi} (t) + \delta \phi (t, {\bf x})$, 
%$A_{\mu} (t, {\bf x}) = \bar{A}_{\mu}(t) + A_{\mu} (t, {\bf x})$ 
in the Robertson-Walker metric
$ds^2 = - dt^2 + a^2(t) d{\bf x}^2$.
We shall study the evolution of the gauge field 
in the Coulomb gauge ${\bf \nabla} \cdot {\bf A} = 0$.
This gauge singles
simply out the dynamical degrees of freedom for the problem
(\ref{scalarlag}) and 
its close relation to the gauge invariant formalism has been
pointed out in \cite{boya}.   
By writing the homogeneous condensate as $\bar{\Phi} (t) = e^{i \Theta(t)}
\rho(t)/\sqrt{2}$ one has at the homogeneous level:
\be
\ddot \rho + 3 H \dot \rho + \frac{\partial
V}{\partial \rho} + e^2 \frac{|\bar{\bf{A}}|^2}{a^2} \rho = 0 
\label{homfield}
\ee   
plus an equation of motion for $\bar{\bf A} (t)$ and the
constraint $\bar{A}_0 = \dot \Theta/e$. We shall consider fluctuations
$\delta A_\mu (t, {\bf x})$ around a background $\bar{A} (t)$ that is
chosen to be zero: 
\begin{eqnarray}
\delta \ddot {\bf A}_T + H \delta \dot {\bf A}_T - \frac{{\nabla}^2}{a^2}
\delta {\bf A}_T 
+ 4 \pi e^2 \rho^2 \delta {\bf A}_T =  4 \pi \delta {\bf j}_T
\label{scalart} \\
- \frac{{\nabla}^2}{a^2} \delta A_0 = 4 \pi \delta j_0 - 4 \pi e^2 \rho^2
\delta A_0 \,,
\label{costraint}
\end{eqnarray}
where $T$ denotes the transverse part of a vector and $\delta {\bf j}_T$
is a source term which can
be different from zero only if one consider quantum or statistical
correlation of the currents \cite{calzetta} because of the
symmetry of the space-time background.
The constraint equation (\ref{costraint}) in Fourier space is 
\be
\left[ \frac{k^2}{a^2} + 4 \pi e^2 \rho^2 \right] \delta A_{0 \, k} = 4
\pi
\delta j_{0 \, k}
\label{constraintk}
\ee
with $\delta j_{0 \, k} = e (\rho \dot{\delta \phi_{2\,k}} - 
\dot \rho \delta \phi_{2\,k} )$ 
where $\delta \phi = (\delta \phi_1 + i \delta
\phi_2)/\sqrt{2}$. 

We consider now the homogeneous part of the differential equation 
(\ref{scalart}) in Fourier space:
\be
\delta {\bf A}''_{T k} + \omega^2_{T k} \delta {\bf A}_{T k} = 0
\label{main}
\ee  
where $\omega^2_{T k} = k^2 + 4 \pi e^2 a^2 \rho^2$ and 
the prime denotes the derivative with respect the conformal 
time $\eta$ ($d \eta = d t/a$). The general solution $y$
of the inhomogeneous differential equation (\ref{scalart})
can be obtained through the Green function method \cite{brand}:
\be
{\bf y} = {\bf y}_h + 4 \pi e^2 \int^\eta \left[ \frac{y_1(\eta')
y_2(\eta) -
y_1(\eta) y_2(\eta')}{W} \right] a^2 {\delta \tilde {\bf j}}_{T k} 
d \eta \prime 
\ee
where $y_h$ is the homogeneous solution, $y_1$, $y_2$ are the
two linear independent homogeneous solutions and $W$ is their wronskian.

Equation (\ref{main}) describes a harmonic oscillator with time
dependent frequency: 
during the oscillation of the complex scalar field Eq. (\ref{main}) can
be 
reduced to a Mathieu-like equation \cite{book}.
The solutions to 
this type of equation show an exponential instability $\propto e^{\mu_k 
\eta}$, for some interval 
of frequencies, called {\em resonance bands}. 
There is a known correspondence at the level of equations of motion 
between Eq. (\ref{main}) and the equation of a massless real scalar field 
$\chi$ (rescaled by its conformal weigth) interacting with a real scalar 
field $\phi$ by a term $g^2 \phi^2 \chi^2$:
\be
(a \chi)''_k + \left[ k^2 + g^2 a^2 \phi^2 + (\xi -\frac{1}{6}) a^2 R
\right]
(a \chi)_k = 0 \,,
\ee
where $ R = 6 a''/a^3 $ is the Ricci curvature. In the case of conformal
coupling ($\xi=1/6$) this correspondence is exact, while for a quartic
potential for $\phi$ (or $\Phi$) this correspondence is {\em nearly} exact 
since $ R \simeq 0$ \cite{greene}.
The behaviour of parametric resonance 
depends strongly on the time behaviour of the homogeneous scalar field, 
which on turn depends on the form of the potential $V(\Phi)$. In the 
following we consider power law potentials as:
\be
V = \lambda_n (\Phi^* \Phi)^{2 n}
\label{potential}
\ee
and we briefly illustrate the
different behaviours depending on 
the parameters $n$ and $\lambda_n$, restricting ourselves to the case in
which the scalar field is the dominant component of energy density in the
universe \cite{coherent}. 

For a quadratic potential ($n=1$) the driving term $a^2 \rho^2$ in Eq.
(\ref{main}) decays as $\eta^{-2}$. 
PR is 
efficient for $4 \pi e^2 \rho^2 \gg \lambda_1$:
the resonance is stochastic and broad, and  
the largest Floquet exponent $\mu_k$ occurs 
for small $k$ fluctuations \cite{klsbig}. 
In the conformally invariant quartic case ($n=2$) the oscillations of 
$\tilde \rho \equiv a \rho$ are given by an elliptic cosine and Eq.
(\ref{main}) reads 
as an equation in Minkowski space-time \cite{greene} (technically speaking
it is a Lame equation). 
The resonance structure and the relative 
Floquet exponents $\mu_k$ depend
non-monotonically by the ratio $e^2/\lambda_2$ 
\cite{greene}: 
long wavelengths $k^2 \ll \lambda_2 \tilde \rho_0^2$ are 
resonant for $1 / 2\pi < e^2/\lambda_2 < 3 / 2\pi$, as well for other
values 
\cite{greene}, where 
$\tilde \rho_0$ is the initial amplitude for $\tilde \rho$. 
For $n \ge 3$ the driving term $a^2 \rho^2$ oscillates and
{\em grows} in time as
$\eta^{\frac{2(n-2)}{2 n - 1}}$. In this case the resonance for $A_{T k}$
(as well for $a \chi$) is very efficient \cite{comment}. 
 
The case of a simmetry breaking potential 
\be
V = m^2 \Phi^* \Phi + \lambda (\Phi^* \Phi)^2 
\label{both}
\ee
with $m^2 < 0$, is definitively the closest to the electroweak phase
transition: 
%the charged scalar field plays the role of the 
$\Phi$ would play the role of the Higgs field, and $\rho$  
rolls down from the zero value (false vacuum) towards 
$\rho_{min} = \pm \sqrt{m^2/ \lambda}$ (true vacuum). 
Let us note that the most important feature in the case of
symmetry breaking in this toy model of scalar electrodynamics is the Higgs
mechanism:
the gauge field gets an effective mass proportional 
to the value of the field in the broken phase $\sqrt{m^2/\lambda}$. 
Besides the unpleasant feature of considering a massive photon, this 
fact could completely inhibit the resonance in an expanding 
universe or narrow the resonance 
shifting it to scales proportional to $\sqrt{m^2/\lambda}$.  
However, in the realistic electroweak theory the photon 
does not acquire a mass: the symmetry
breaking mechanism gives masses to the $W$ and $Z$ bosons, leaving 
the photon massless. This is an important 
condition for the efficiency of parametric resonance in expanding
universe. 
We note that in this case the driving term in Eq. (\ref{main}) is given by  
the $W$ condensate \cite{new}.

Let us discuss for completeness also the behaviour of the scalar 
field fluctuations for the potential (\ref{both}) without any constraints on
the sign of $m^2$. The Fourier components of the real part of the rescaled
field fluctuation $\widetilde{\delta \phi_1} \equiv a \delta \phi_1$ satisfy
the following equation:
\be
\widetilde{\delta \phi}''_{1\,k} + 
%3 H \dot{\delta \phi}_{1\,k} +
\left[ \omega^2_{1\,k} - \frac{a''}{a} \right]
\widetilde{\delta \phi}_{1\,k} = 0
\label{real}
\ee
\noindent
where $\omega^2_{1\,k} = k^2 + a^2 m^2 + 3 \lambda a^2 \rho^2$.
The Fourier
components of the imaginary part of the rescaled field fluctuation
$\widetilde{\delta
\phi_2} \equiv a \delta \phi_2$ satisfy
\begin{eqnarray}
\widetilde{\delta \phi}''_{2\,k} +
\left[ \omega^2_{2\,k} - \frac{a''}{a} \right]
\widetilde{\delta \phi}_{2\,k} 
= \nonumber \\ e a^2 (2 \delta A_{0\,k} \rho' + \rho \delta A'_{0\,k} + 3
a H \rho \delta A_{0\,k})
\label{imaginary}
\end{eqnarray}
where $\omega^2_{2\,k} = k^2 + m^2 a^2 + \lambda a^2 \rho^2$.
By using the constraint
equation (\ref{constraintk}) and introducing $q_k = \widetilde{\delta
\phi}_{2\,k} / \omega_T$ one obtains  
%\be
\begin{eqnarray}
q_k'' +
\left[ \omega^2_{k\,2} + \omega^2_T - k^2 - \frac{a''}{a} + \frac{8 \pi
e^2 a^2}{\omega^2_T} (\rho' + a H \rho)^2 \right. \nonumber \\ \left.
+ \frac{\omega_T''}{\omega_T} - 2 \frac{\omega_T'^2}{\omega_T^2} \right]
q_k = 0 \,.
\label{imaginary2}
\end{eqnarray}
\noindent
We note that the gauge coupling affects the resonance structure 
of the scalar field: while $\omega_{1 k}^2$ is the driving term 
for $\widetilde{\delta \phi}_{1 k}$, 
$\omega_{2 k}^2$ does not determine the resonance bands for
$\widetilde{\delta
\phi}_{2 k}$, as would do if the charged scalar field were uncoupled to a
gauge field \cite{note}. 

\section{Axion Electrodynamics} 

We consider the effect of parametric
resonance for another type of coupling described by:
%of electromagnetism to a scalar field: 
\be 
{\cal L} = - \frac{1}{16\pi} F_{\mu \nu} F^{\mu \nu} -
\frac{1}{2} \partial_\mu \phi   
\partial^\mu \phi - V(\phi) - \frac{g}{4} \phi F_{\mu \nu} 
{\tilde F}^{\mu \nu}
\label{pseudo}
\ee
The neutral scalar field $\phi$ can  represent an axion in \cite{turner} 
(in such a case the vacuum angle $\theta$ is defined to be $\theta =
\phi/f$, where $f$ is the Peccei-Quinn simmetry scale)
or a general pseudo-Goldstone boson in \cite{garretson}. 
By considering again the scalar field as $\phi(t, {\bf x}) = \phi(t) +
\delta \phi (t, {\bf x})$ one has at the homogeneous level:
\be 
\ddot \phi + 3 H \dot \phi + \frac{\partial
V}{\partial \phi} = 0
\label{homeqaxion}
\ee 
and for the transverse degrees of freedom 
\be
\delta \ddot {\bf A}_T + H \delta \dot {\bf A}_T - \frac{{\nabla}^2}{a^2}
\delta {\bf A}_T
+ 4 \pi g \dot \phi \frac{\bf \nabla}{a} \times \delta {\bf A}_T = 0 \,.
\label{axiont}
\ee 
In this case it is more convenient to consider the circular polarized
Fourier
component perpendicular to ${\bf k}$, ${\bf A}_{\pm\,k}$, whose equation
of motion
is  
\be 
\delta {\bf A}''_{\pm \, k} + ( k^2 \pm 4 \pi g \phi ' k) \delta 
{\bf A}_{\pm\,k} = 0
\label{axion}
\ee
In the regime of coherent oscillation of the scalar field 
one has again a Mathieu-like equation where the driving term is
proportional to the time derivative of the scalar field and goes as
$\eta^{-1}$ for a power law potential analogous to 
Eq. (\ref{potential}) (always under the assumption that the scalar field
dominates the energy density of the universe \cite{coherent}). 
The efficiency of the resonance depends on the value of the adimensional
quantity $g \phi$. 
%where $f_a$ is the scale of the scalar field.
The analysis of Eq. (\ref{axion}), describing the axion decay into
photons, was probably one of the first application 
of parametric resonance in cosmology
\cite{first}, and it has been recently reanalyzed in
\cite{LeeNg,brustein}. 
By considering $\phi$ as the QCD axion, the smallness of $g f
\sim 10^{-4}$ prevents the efficiency of the resonance in the case 
of a potential with temperature-dependent mass due to QCD effects
\cite{LeeNg}. 
In general the resonance is on $k \sim 4 \pi g \phi \, \omega$, where 
$\omega$ is the oscillation frequency of $\phi$.
However, for $V = \lambda \phi^4/4$ and $4 \pi g f=1$ we observe a
linear
growth of $A_{\pm}$ for $k/\omega \ll 4 \pi g f$ 
by numerically integrating Eq. (\ref{homeqaxion}) and (\ref{axion}). Such
growth is slower than the quasi-exponential amplification of the field
fluctuations $\delta \phi$ on smaller scales \cite{greene}. 

Besides the PR effect, one of the two circular polarization exhibit 
also a negative
effective mass (depending on the sign of $\dot \phi$). This feature could
be interesting before the stage during which the field coherently
oscillates, when $\dot \phi$ has a fixed sign. 

\section{Plasma effects} 
We now comment on effects due to the presence of
other charged particles \footnote{We are referring to charged fields which
are uncoupled from the scalar field which oscillates coherently. These
fields produce currents $J_\mu$ which
are {\em incoherent} and are coupled in a $J_\mu A^\mu$ to the gauge
field.}. These effects can be
important since the universe is considered a good conductor after
reheating until decoupling. For wavelengths larger than 
the collision length \cite{jackson} one could use the Ohm relation 
for the additional current ${\bf j}_{\rm add} = \sigma {\bf E}$.
This would add to Eqs. (\ref{main}) and (\ref{axiont}) a damping term:
\be
\delta {\bf A}''_k + 4 \pi (a \sigma)
\delta {\bf A}'_k + \omega^2_k (\eta) \delta {\bf A}_k = 0 \,,
\label{sfiga}
\ee
where we have denoted collectively the transverse component as $\delta
{\bf A}$ and
with $\omega^2_k$ their time dependent frequency.
The time behaviour of conductivity is considered in the literature
to be $\sigma
\propto T \propto 1/a$
\cite{turner} and so the term multiplying the first
derivative in Eq. (\ref{sfiga}) is constant: in this way
the resonance on long wavelengths could be  
washed out if $a \sigma$ is larger than the Floquet exponent. For
wavelengths smaller than the collision length charge
separation should be taken into account and a term qualitatively 
similar to the classical plasma frequency would enter as:
\be
\omega^2_k (\eta) \rightarrow \omega^2_k (\eta) + 4 \pi e^2
\frac{n(\eta)}{m}
\label{plasma}
\ee
where $n(\eta)$ and $m$ are respectively the number density and the mass 
of the charged particles. This term would play the role of an effective
mass, but which would decay as $a(t)^{-3}$ (which is more rapid than the 
decay of the driving term considered in this paper),
leaving open the possibility
of resonance, in particular for large coupling costants. 
Let us observe that these qualitative estimates for conductivity and 
plasma frequency 
are obtained by thermodynamic considerations. Even in the case of a
preexisting 
thermal equilibrium the resonance would 
%push GF fluctuations out of
%equilibrium 
amplify the gauge fluctuations. Infact a
thermal mass $\sim T^2$ for the gauge fiels will not be able to prevent
the resonance and it has the effect of shifting the resonance on smaller
scales ($k^2$ must be replaced with $k^2 + g_{\rm eff} T^2 a^2$). 

There is an epoch where the the resonance 
could proceed just as worked out in the vacuum case: in preheating after
inflation, when it is assumed that the "creation" of charged particles is 
contemporaneous to the amplification of gauge fluctuations. This
amplification will last until rescattering and
backreaction set in to shut off the resonance \cite{serguei}, and terms
such as
conductivity and plasma
frequency will appear in this out-of-equilibrium process 
by a self-consistent study
of the problem
\cite{simio}(although their time-dependence will be different from 
the corresponding thermal equilibrium quantities).
In this case the constraints on the coupling costants $e^2$ and $g$ 
come only from requirement to protect the 
potential of the scalar field from radiative corrections: 
this argument would give the usual $e^2 \lesssim 10^{-6}$ 
for the scalar electrodynamic case. If one requires that the 
charged inflaton gives the right amount of e-folds and CMB fluctuations 
($m \sim 10^{-6} M_{pl}$ for a massive inflaton, $\lambda \sim 10^{-13}$ 
for a self-interacting inflaton), then a broad resonance
regime would be allowed for the gauge field. 
For the axion electrodynamic case, the same argument
leads to $g f \gg 10^{-4}$, opening a possible new efficient channel for
the decay of the scalar field $\phi$ into gauge fluctuations.
 
\section{Primordial Magnetic Field by Parametric Resonance?} 
Even if the conformal property of gauge fields in cosmological
backgrounds is useful for PR, it is the main reason why
gauge fluctuations are not
amplifield by the expansion of the universe \cite{conformal}, in
contrast to what happens for minimally coupled scalar fields.
Therefor the conformal invariance of gauge fields is the main issue for
the explanation of primordial magnetic fields \cite{dynamo} within the
inflationary paradigm \cite{turner,garretson}.

PR could be interesting for the generation of primordial magnetic fields
for two reasons.
First, it is intringuing to note how the exponential growth due to
parametric
resonance is of the same type as the dynamo effect \cite{dynamo}, one of
the
astrophysical processes postulated to explain the observational evidence
for galactic magnetic fields. 
As a second point, there are several examples in which
significant
amplification occurs on the
maximum causal scale allowed by the problem (the coherence scale of the
field).
Since the scale of the magnetic seeds is always a crucial issue for a
model which aims to explain their origin, this feature of PR is
very interesting. 

As a straightforward result of section II preheating could provide an
extra growth in the amplitude of gauge fluctuations 
which are produced during inflation
\cite{turner,garretson}: for parameters in which long wavelength gauge 
fluctuations are amplified, this occurs directly on observable scales
because of the coherence of the inflaton on the particle horizon scale. 

We illustrate how the mechanism could work in a simple scalar electrodynamic 
toy model where the charged scalar field is a massless 
self-interacting inflaton 
with $2 \pi e^2 = \lambda$ 
%(with $\lambda \sim 10^{-13}$ 
%in order to have
%density fluctuations in agreement with the anisotropies of the 
%comoving background radiation). 
During inflation, assuming $\Phi$ in slow rollover, 
a solution for Eq. (\ref{main}) 
%which matches an adiabatic vacuum in the infinite past 
is:
\be 
\delta {\bf A}_{T \, k} = (- H \eta)^{1/2} \left( \frac{\pi}{4 H}
\right)^{1/2} 
H_\nu^{(1)} (- k \eta)
\label{solution}
\ee
where $\eta = - 1/(H a)$ is the conformal time used to model a de Sitter era 
($-\infty < \eta < \eta_0 < 0$, where $\eta_0$ is some arbitrary time) and 
the index of the Hankel function is so defined
\be 
\nu^2 = \frac{1}{4} - 4 \pi e^2 \frac{\rho^2}{H^2} \simeq \frac{1}{4} -
\frac{3 e^2}{2 \lambda} 
\frac{M^2_{\rm pl}}{\rho^2} 
\ee
where the last equality holds during slow-rollover and $M^2_{\rm pl}$ is
the inverse of the Newton constant $G$.
The solution (\ref{solution}) matches the 
adiabatic vacuum $e^{- i k \eta}/\sqrt{2 k}$ for $\eta \rightarrow - \infty$. 
For $\nu^2 > 0$ Eq. (\ref{solution}) leads 
%at most to a slightly red 
to a slight shift 
of the initial vacuum for the long wavelength fluctuations 
($-k \eta \rightarrow 0$)
of the gauge field:
\be
\delta {\bf A}_{T \, k} \simeq - i \frac{\Gamma (\nu)}{\sqrt{2 \pi k}} 
\left( \frac{- k \eta}{2} \right )^{\frac{1}{2} - \nu} \,.
\label{long}
\ee
By using the definition $B_i = \epsilon_{i m n} \partial_m A_n / a$ 
one can relate 
the fluctuation of the magnetic field to the gauge field fluctuations 
%and 
%define the ratio $r$ of the energy stored in the magnetic field with
%respect to the total one:
\be
|{\bf B}_k|^2 = \frac{k^2 |\delta {\bf A}_{T \, k}|^2}{a^4}
%\,, \quad   
%r \equiv \frac{\rho_B}{\rho_{\rm tot}}
\label{definitions}
\ee
Parametric resonance after inflation amplifies the gauge fluctuations in 
Eq. (\ref{long}) and $|{\bf B}_k|^2$ in Eq. (\ref{definitions}) 
exponentially up to a maximum factor $10^{12}$ \cite{serguei}. 

However, this last amplification is not 
sufficient to explain the value required by the observation. Indeed, 
the simple toy model of a self-interacting inflaton is very useful 
in order to follow gauge fluctuations at a linear order from 
inflation through preheating, but it misses a super-adiabatic amplification 
during inflation, which was investigated by Turner and Widrow
\cite{turner}. This
amplification can occur due to an 
effective {\em negative} mass for the gauge field during inflation,
which can be obtained by breaking the conformal invariance
through a explicit coupling to the curvature, for instance \cite{turner}. 
In string cosmology an effective negative mass for the gauge field is
generated by the dilaton coupling \cite{string}. 
%or due to a
%dilaton coupling in string cosmology \cite{string}. 

Instead, as we see from Eq. (\ref{long}), 
long wavelength gauge fluctuations are slightly modified in amplitude
and spectrum with respect to the initial adiabatic spectrum by the coupling
with inflaton during slow-rollover. Indeed the {\em positive} mass
generated by the coupling of the gauge field to the charged scalar field
lead to a slight suppression.
When slow-rollover is ended and $\nu^2 < 0$ the gauge modes are
suppressed. This phase is just a
short transient, after which the resonant amplification starts
immediately (roughly when $\rho$ first cross zero).  
The main point is the decay of $B^2$ as $1/a^4$ once a gauge
fluctuation has left the Hubble radius during inflation, leading to a
suppression which the following preheating phase cannot fill up
\cite{lwsupp}.

Now we quantify our previous statements and we estimate the order of
magnitude of the amplitude for the magnetic
field at decoupling. We shall consider the effect of
preheating as a k-independent amplification factor $\Omega^2$ which
multiplies the spectrum at the end of inflation (\ref{long}). We
also neglect the finite time duration of the preheating phase and we
identify the end of inflation with the beginning of the usual
radiation dominated Friedmann era.
At the beginning of the radiation dominated phase (denoted in the
following by $\eta^*$) the spectrum of the magnetic field is:
\be
k^3 |{\bf B}_k (\eta^*)|^2 = \Omega^2 \frac{k^5}{a^4 (\eta^*)} 
|\delta {\bf A}_{T\,k} (\eta^*)|^2
\label{estimate1}
\ee
where we consider $\delta {\bf A}_{T\,k} (\eta^*) \simeq 1/\sqrt{2 k}$,
which has
been obtained from Eq. (\ref{long}) by setting $\nu = 1/2$. By using the
the relation $a^4 B^2 = const.$ for the magnetic field evolution 
in the radiation dominated era, the spectrum at the
decoupling time $\eta_{\rm DEC}$ is:
\be
k^3 |{\bf B}_k (\eta_{\rm DEC})|^2 = k^3 |{\bf B}_k (\eta^*)|^2
\frac{T_{\rm DEC}^4}{T_{\rm REH}^4}
\label{estimate2}
\ee
since the temperature $T$ scales as the inverse of the scale factor $a$.
By using Eqs. ({\ref{estimate1}) and (\ref{estimate2}) we estimate at time 
of decoupling a magnetic field of size 
\begin{eqnarray}
{\cal B} |_{\rm DEC} (k) &=& ( k^3 |{\bf B}_k (\eta_{\rm DEC})|^2
)^{1/2} \nonumber \\
&\simeq& \frac{\Omega}{\sqrt{2}} \, ( k {\rm kpc} )^2 \frac{T_{\rm
DEC}^2}{ {\rm kpc}^2 T_{\rm NOW}^2} \nonumber \\ 
&\simeq& 10^{-42}\, {\rm Gauss}
\label{estimate}
\end{eqnarray}
where it has been assumed $a(\eta_{\rm NOW}) = 1$, a comoving wavenumber
$k = 0.1 \, {\rm kpc}^{-1} = 0.64 \,\times\,  10^{-36} {\rm GeV}$,
$\Omega = 10^6$, $T_{\rm DEC}/T_{\rm NOW} \simeq 10^{3}$, 
$1 \, {\rm Gauss} = 3 \times 10^{-20} {\rm GeV}^2$ \cite{kolb}.
The estimate (\ref{estimate}) is lower than the magnetic seed field
required by observation, which is $10^{-34}$G for a flat, low density
universe \cite{davis}.
We note that ${\cal B}$ is proportional 
to $\Omega$, but not to $T_{\rm REH}$ or to the Hubble constant
during inflation, as observed in \cite{turner}. This is due to the the
approximation $\nu = 1/2$ which restablishes the conformal invariant
spectrum in the estimate (\ref{estimate}).

\section{Conclusions} 
In this paper we have studied the possibility of 
an enhancement of gauge fields due to parametric resonance. 
%in the models of scalar and axion electrodynamics in an expanding
%universe. 
For both the couplings to a scalar charged field and to an axion, the
resonance can counteract the redshift due to the expansion of the universe 
and exponential growing modes appear. The resonance is generically 
on small scales 
%comparable to 
%the effective mass for the gauge field 
in the axion case, while it can occur 
for scales up to the coherence scale of the scalar field in the scalar 
electrodynamic case. Therefore the latter scalar field coupling 
is more interesting for the problem of cosmological magnetic fields, 
in which the large scale observed seems a crucial issue. 

In the simplest toy model with a massless charged inflaton, 
a linear calculation based on preheating does not provide directly an
amplitude in agreement with observations because 
the magnetic field fluctuations are redshifted during inflation. 
However, this latter effect may be circumvented in many ways.
In more realistic inflationary models motivated by particle physics
the gauge field can be sustained during inflation 
and then amplified during preheating.
A sustain during inflation could be also realized by breaking
conformal invariance with explicit coupling between the gauge field 
and the Riemann curvature in the lagrangian \cite{turner}. Conformal
invariance is
also effectively broken by the inhomogeneities genarated during inflation
(a second order effect in perturbations).
Moreover, PR would also provide an additional growth in the generation
mechanism based on 
the statistical correlation of the currents studied in \cite{calzetta}.

A final amplitude and spectrum of gauge field fluctuations generated 
by PR is however beyond the scope of the present paper. Infact,  
its calculation would include the analysis 
of nonlinear effects and a simple estimate based on matching techniques 
is inadequate. In models with interacting scalar fields, nonlinear 
effects as rescattering \cite{serguei} play an important role in the 
generation of the final power spectra. Indeed, such power spectra, 
as $|\delta {\bf A}_k|^2 \sim k^{-\alpha}$ with $\alpha > 1$, 
would be very interesting for the problem of large scale magnetic
fields, because steeper than (\ref{long}) in the infrared region. 

\vspace{1cm}

{\bf Acknowledgments}

\noindent
It is a pleasure to thank R. Brandenberger for discussions and
many valuable comments on earlier versions of this draft, and S.
Khlebnikov and G. Venturi for discussions and suggestions. 
We thank R. Balbinot, B. Bassett, K. Dimopoulos, K.
Enqvist, G. Field, E. Lazzaro, A. Riotto, A. Starobinsky, K. Subramanian, 
O. Tornkvist, G. Veneziano for discussions and comments. F. F. was
supported in part 
by the DOE under Grant no. DE-FG02-91ER40681 (Task B).


\begin{thebibliography}{10}

\bibitem{tutti} J. Traschen and R. Brandenberger, {\em Phys. Rev.}
D {\bf 42}, 2491 (1990); L. Kofman, A. Linde, and A. Starobinsky,
{\it Phys. Rev. Lett.} {\bf 73}, 3195 (1994); Y. Shtanov,
J. Traschen, and R. Brandenberger, {\em Phys. Rev.} D {\bf 42}, 2491 (1995); 
M. Yoshimura, {\em Prog. Theor. Phys.} {\bf 94}, 873 (1995); D.
Boyanovsky, H. de Vega, 
and R. Holman, hep-ph/9701304, and references therein; D. Kaiser, {\em
Phys. Rev.} D {\bf 53}, 1776 (1996).  
%
\bibitem{fermion} J. Baacke, K. Heitmann, and C. Patzold, {\em Phys. Rev.} 
D {\bf 58}, 125013 (1998);
P. Greene and L. Kofman, {\em Phys. Lett.} {\bf B 448}, 6 (1999).  
%
\bibitem{gw} S. Khlebnikov and I. Tkachev, {\em Phys. Rev.} D {\bf 56},
653 (1997); B. Bassett, {\em Phys. Rev.} D {\bf 56}, 3439
(1997).
%
\bibitem{scalar} B. Bassett, D. Kaiser, and R. Maartens, {\em Phys. Lett.}
B {\bf 455}, 84 (1999); F. Finelli and R. Brandenberger, 
{\em Phys. Rev. Lett.} {\bf 82}, 1362 (1999); M. Parry and
R. Easther, {\em Phys. Rev.} D {\bf 59}, 061301 (1999).
%
\bibitem{gravit} A. Maroto and A. Mazumdar, hep-ph/9904206; 
R. Kallosh, L. Kofman, A. Linde, and A. Van Proeyen,
hep-ph/9907124; G. F. Giudice, I. Tkachev, and A. Riotto,
JHEP {\bf 9908}, 009 (1999).
%
\bibitem{nearly} By the term {\em nearly} we do mean not
only fields whose equation of motion are {\em exactly} conformally
related
to the Minkowski case, but also approximatively (as for instance massless
scalar fields in a radiation dominated universe \cite{greene}).
%
\bibitem{greene} P. Greene, L. Kofman, A. Linde, and A. Starobinsky,
{\it Phys. Rev.} {\bf D56}, 6175 (1997).
%
\bibitem{klsbig} L. Kofman, A. Linde, and A. Starobinsky,
{\it Phys. Rev.} D {\bf 56}, 3258 (1997); S. Khlebnikov and I. Tkachev,
{\em Phys. Lett.} B {\bf 390}, 80 (1997).
%
%\bibitem{wide} S. Khlebnikov and I. Tkachev, {\em Phys. Lett.}
%B {\bf 390}, 80 (1997). 
%
\bibitem{conformal} L. Parker, {\em Phys. Rev.} D {\bf 3}, 346 (1971) and 
{\em Phys. Rev.} D {\bf 5}, 2905 (1972); 
F. Finelli, A. Gruppuso and G. Venturi, {\em Class. Quant. Grav.}
{\bf 16}, 3923 (1999). 
%
%\bibitem{giovannini} The resonant production of photon due to the coupling with
%a dilaton  
%in string cosmology has been analyzed by M. Giovannini, {\em Phys. Rev.} D
%{\bf 56}, 631 (1997). 
%
\bibitem{boya} D. Boyanovsky, D. Brahm, R. Holman, and D.-S. Lee,
{\em Phys. Rev.} D {\bf 54}, 1763 (1996).   
%
\bibitem{calzetta} E. Calzetta, A. Kandus, and F. Mazzitelli, {\em
Phys. Rev.} D {\bf 57}, 7139 (1998); M. Giovannini and M. Shaposhnikov, 
Phys. Rev. D {\bf 62}, 103512 (2000). 
%
%\bibitem{new} F. Finelli, A. Gruppuso, in preparation.
%
\bibitem{brand} V. Zanchin, A. Maia, Jr., W. Craig, and R. Brandenberger,
{\em Phys. Rev.} D {\bf 57}, 4651 (1998).
%
\bibitem{book} N. W. Mac Lachlan, {\em Theory and Application of Mathieu
functions}, (Dover, 1961).
%
\bibitem{coherent} M. Turner, {\em Phys. Rev.} D {\bf 28}, 1243 (1983); 
Y. Nambu, and A. Taruya, {\em Prog. Theor. Phys.} {\bf 97}, 83 (1997).
%
\bibitem{comment} We note that that the rescaled fluctuations of the field 
itself $a \delta \phi$ would be driven by $a^2 V''$, whose behaviour is 
$\eta^{\frac{4 (2 - n)}{2 n -1}}$ \cite{coherent} and therefore it decays
in time for $n \ge 3$.
%
\bibitem{new} F. Finelli, A. Gruppuso, in preparation.
% 
\bibitem{note} The study of the resonance for $\delta \phi_2$ is not
trivial as it can seems. In both the equations for $\delta \phi_{2\,k}$
and $q_k$ there are singularities when $w_T = 0$, which occurs when
$\rho = 0$ for the interesting case of long wavelength modes.
%
\bibitem{turner} M. Turner and L. Widrow, {\em Phys. Rev.} D {\bf 37},
2743 (1988). 
%
\bibitem{garretson} W. Garretson, G. Field, S. Carroll, 
{\em Phys. Rev.} D {\bf 46}, 5346 (1992).  
%
%\bibitem{jackson} J. D. Jackson, {\em Classical Electrodynamics}, 
%Wiley (1975).
%
\bibitem{first} J. Preskill, M. Wise, and F. Wilczeck, {\em Phys.
Lett.} B {\bf 120}, 127 (1983); L. Abbott and P. Sikivie, {\em ibid.},
{\bf 120}, 133 (1983).   
%
\bibitem{LeeNg} D.-S. Lee and K.-W. Ng, Phys. Rev. {\bf D 61}, 085003
(2000) 
%
\bibitem{brustein} R. Brustein and D. Oaknin, {\em Phys. Rev. Lett.} {\bf
82}, 2628 (1999).
%
\bibitem{jackson} J. D. Jackson, {\em Classical Electrodynamics},
Wiley (1975).
%
\bibitem{serguei} S. Khlebnikov and I. Tkachev,
{\em Phys. Rev. Lett.} {\bf 79}, 1607 (1997). 
%
\bibitem{simio} D. Boyanovsky, H. J. de Vega, and M. Simionato,
Phys. Rev. {\bf D 61}, 085007 (2000).
%
\bibitem{dynamo} E. N. Parker, {\em Cosmical Magnetic Fields} (Clarendon,
Oxford, England, 1979); Y. Zel'dovich, A. Ruzmaikin, and D. Sokoloff, {\em
Magnetic Fields in Astrophysics} (Gordon and Breach, New York, 1983)
%
\bibitem{string} B. Ratra, {\em Ap. J. Lett.} {\bf 391}, L1 (1992); 
M. Gasperini, M. Giovannini, and G. Veneziano, {\em Phys. Rev. Lett.} {\bf
75}, 3796 (1995).
%An effective negative mass for the gauge field is also
%generated by the dilaton coupling in string cosmology: M. Gasperini, M.
%Giovannini, and G. Veneziano, {\em Phys. Rev. Lett.} {\bf 75}, 3796
%(1995).
%
%\bibitem{dolgov} A. Dolgov, {\em Phys. Rev.} D {\bf 48}, 2499 (1993);
%F. Mazzitelli
\bibitem{lwsupp} A similar effect occurs in the scalar sector for
a scalar field $\chi$ which is strongly coupled to a massive inflaton.
The large effective mass of $\chi$ during inflation leads to a suppression
for the long wavelength $\chi$ fluctuations:  K. Jedamzik and G. Sigl,
{\it Phys. Rev.} {\bf D91}, 023519 (2000); P. Ivanov, {\it Phys. Rev.}
{\bf D61}, 023505 (2000).
%
\bibitem{kolb} E. W. Kolb and M. S. Turner, {\em The Early Universe},
(Addison-Wesley, Redwood City, California, 1990).
%
\bibitem{davis} A.-C. Davis, M. Lilley, and O. Tornkvist, 
{\it Phys. Rev.} {\bf D 60}, 021301 (1999).
%
\end{thebibliography}
\end{document}